# Machine Learning-based Anomaly Detection in Optical Fiber Monitoring

Khouloud Abdelli,[1,3] * Joo Yeon Cho,[1] Florian Azendorf,[2] Helmut Griesser,[1] Carsten Tropschug,[2] Stephan Pachnicke[3]

[1]ADVA Optical Networking SE, Fraunhoferstr. 9a, 82152 Munich/Martinsried, Germany
[2]ADVA Optical Networking SE, Märzenquelle 1-3, 98617 Meiningen, Germany
[3] Christian-Albrechts-Universität zu Kiel, Kaiserstr. 2, 24143 Kiel, Germany
*Corresponding author: kabdelli@adva.com

**Secure and reliable data communication in optical networks is critical for high-speed Internet. However, optical fibers, serving as the data transmission medium providing connectivity to billons of users worldwide, are prone to a variety of anomalies resulting from hard failures (e.g., fiber cuts) and malicious physical attacks (e.g., optical eavesdropping (fiber tapping)) etc. Such anomalies may cause network disruption and thereby inducing huge financial and data losses, or compromise the confidentiality of optical networks by gaining unauthorized access to the carried data, or gradually degrade the network operations. Therefore, it is highly required to implement efficient anomaly detection, diagnosis, and localization schemes for enhancing the availability and reliability of optical networks. In this paper, we propose a data driven approach to accurately and quickly detect, diagnose, and localize fiber anomalies including fiber cuts, and optical eavesdropping attacks. The proposed method combines an autoencoder-based anomaly detection and an attention-based bidirectional gated recurrent unit algorithm, whereby the former is used for fault detection and the latter is adopted for fault diagnosis and localization once an anomaly is detected by the autoencoder. We verify the efficiency of our proposed approach by experiments under various anomaly scenarios using real operational data. The experimental results demonstrate that: (i) the autoencoder detects any fiber fault or anomaly with an F1 score of 96.86%; and (ii) the attention-based bidirectional gated recurrent unit algorithm identifies the the detected anomalies with an average accuracy of 98.2%, and localizes the faults with an average root mean square error of 0.19 m.**

## 1. INTRODUCTION

Optical fiber is the essential medium for transporting a large amount of data through the aggregated Internet, mobile backhaul and core network. A single fiber link connects thousands of customers and enterprises, carrying a mixture of personal, business, and public data. Therefore, the impact of a broken fiber can be enormous and must be responded to immediately.

In general, optical fiber is vulnerable to different types of anomalies including fiber cut, fiber eavesdropping (fiber tapping) etc. Such anomalies compromise the availability and the confidentiality of an optical network. Specifically, the manual discovery of incidents occurring in the fiber requires considerable expert knowledge and probing time until a fault (e.g., broken fiber) is identified. Fiber monitoring aims at detecting anomalies in an optical layer by logging and analyzing the monitoring data. It has mainly been performed using optical time domain reflectometry (OTDR), a technique based on Rayleigh backscattering, widely applied for fiber characteristics' measurements and for fiber fault detection and localization [1]. OTDR operates like an optical radar. It sends a series of optical pulses into the fiber under the test. The backscattered signals are then recorded as a function of time, which can be converted into the position on the optical fiber. As result, a recorded OTDR trace illustrating the positions of faults along the fiber, is generated, and used for event analysis. However, OTDR traces are difficult to interpret even by highly experienced engineers mainly due to the noise overwhelming the signals. Analyzing OTDR signals using conventional methods can be time consuming as performing a lot of averaging of OTDR measurements is required to remove the noise and thereby to achieve a good event detection and localization accuracy. Therefore, it would be highly beneficial to develop a reliable automated diagnostic method that accurately and quickly detects, diagnoses and pinpoints fiber faults given the OTDR data and thereby reducing operation-and-maintenance expenses (OPEX) and eliminating the time needed to investigate the cause and determine a search area. Upon finding the fault location, appropriate action is taken to remedy the fault and restore service as quickly as possible.

Recently, machine learning (ML)-based approaches have shown great potential to tackle the problem of fiber event detection and localization [2]. In this respect, long short-term memory, and convolutional neural networks (CNNs) have been proposed to detect and localize the reflective fiber events induced by the connectors and mechanical splices [3-5]. A hybrid ML-based framework combining a bidirectional long short memory (BiLSTM) network and CNNs, called BiLSTM-CNN, has been presented for detecting, localizing, and discriminating between reflective, non-reflective and merged events [6]. To tackle the challenge of fiber event analysis under very low conditions (SNR levels ranging from -5 dB and 15 dB), a method combining a denoising convolutional autoencoder (DCAE) and a BiLSTM has been proposed, whereby the former is used for noise removal of OTDR signals and the latter, trained with noiseless OTDR signals, is adopted for fault detection, localization, and diagnosis with the denoised signal by DCAE as input [7]. The ML models presented in the last two publications were trained using experimental data incorporating faults modeled using optical components such as connectors or reflectors. Therefore, the generalization and robustness capabilities of such models may severely degrade when tested with new unseen data including real induced faults with various patterns such as fiber bend events generated for different bending radius values. Furthermore, although such methods distinguish the non-reflective events from the other events, they cannot easily discriminate the faults due to bad splice or fiber tapping. Generally, the generated fiber fault data for training ML-based diagnostic methods is highly unbalanced (the instances of normal states are higher than the faulty instances), which inevitably raises the issues caused by unbalanced class distributions. Additionally, it can be prohibitively expensive and cumbersome to obtain data representing all types of faults or anomalies generated for different scenarios or settings that are accurately labeled. That is why unsupervised ML techniques, particularly reconstruction-based anomaly detection approaches, are frequently adopted.

In this paper, an unsupervised ML technique, an autoencoder, is proposed to quickly detect any anomaly or unexpected abnormal event patterns in optical fibers. Once an anomaly or fault is detected, a diagnostic ML model adopting attention mechanisms and bidirectional gated recurrent unit network, is used to diagnose and localize the fault. The proposed methods are applied to noisy OTDR data with SNR levels varying from 0 dB to 30 dB, including several real faults induced at different locations of an optical network. Our contributions can be summarized as follows:

- An autoencoder based anomaly detection model is proposed for detecting any faults in fiber optics including fiber cut and fiber tapping attack.
- An attention-based bidirectional gated recurrent unit model for fiber fault diagnosis and localization is presented.
- The efficiency of the proposed methods is validated using experimental monitoring data.

Throughout this article "Diagnose" refers to the operation of discriminating between the different types of faults. "Localize" stands for the act of estimating the location of the fault. "Detect" refers to the operation of finding out any abnormal behavior in optical fiber.

The rest of this paper is structured as follows: Section 2 gives some background information about the physical fiber attacks, the autoencoder, and bidirectional gated recurrent unit algorithm. Section 3 presents the proposed framework for predictive fiber monitoring. Section 4 describes the experimental setup and the validation of the presented approach. Conclusions are drawn in Section 5.

## 2. BACKGROUND
### 2.1 Fiber anomalies

### 2.1.1 Fiber cut
Fiber cut is a disastrous physical failure in optical networks, capable of causing widespread disruption. In most cases fiber cuts are the result of accidental cable damages due to construction activities, ship anchors at cable landing points, or natural disasters like earthquakes or tornadoes, and only rarely by intentional damage such as sabotage with the malevolent intent to induce a denial-of-service. Fiber cuts are considered as the single largest cause of service outages. As reported by the Federal Communication Commission (FCC), more than one-third of service disruptions are caused by fiber-cable breaks [8]. Any service outage due to a fiber cut results in massive data loss, network disruption, and huge financial loss etc [9]. In 1991, a severed fiber-optic cable shut down all New York airports and induced air traffic control problems [10]. It is time-consuming to locate and repair fiber cuts.

### 2.1.1 Optical eavesdropping
Optical eavesdropping attack permits the eavesdropper to gain an unauthorized access to the carried data by directly accessing the optical channel via fiber tapping for the purpose of stealing mission-critical and sensitive information. There are several fiber tapping techniques which can be adopted to launch the eavesdropping attack, such as fiber bending, optical splitting, evanescent coupling, V Groove cut etc [11]. However, the easiest method to make the eavesdropping intrusion undetected is microfiber bending by using commercially available clip-on coupler. Fiber-optic cable tapping incidents have been reported such as the eavesdropping device, which was discovered illegally installed on Verizon's optical network in 2003 to glean information from a mutual fund company regarding its quarterly statement prior to its public disclosure [12]. Although, it is easy to perform an eavesdropping intrusion, it is challenging to detect such intrusion using conventional intrusion detection methods such as OTDR-based techniques.

### 2.2 Autoencoder
An autoencoder (AE) is a type of artificial neural network seeking to learn a compressed representation of an input in an unsupervised manner [13]. As shown in Fig. 1, an AE is composed of two sub-models namely the encoder $f_\theta$ and the decoder $g_\theta$. Generally, the encoder and the decoder are of symmetrical architecture comprising of several layers each succeeded by a nonlinear function and shared parameters $\theta$. The encoder $f_\theta(.)$ compresses an input $\boldsymbol{x}$ into lower-dimensional latent-space representation $\boldsymbol{z}$, and the decoder $g_\theta(.)$ maps the encoded representation back into the estimated vector $\hat{\boldsymbol{x}}$ of the original input vector. For an autoencoder composed of multiple layers, the nonlinear encoder and decoder mappings can be formulated as follows:

$$f_\theta^l(.) = \sigma^l \left( \boldsymbol{W}^l \left( f_\theta^{l-1}(.) \right) + \boldsymbol{b}^l \right) \quad (1)$$
$$g_\theta^l(.) = \sigma'^l \left( \boldsymbol{W'}^l \left( g_\theta^{l-1}(.) \right) + \boldsymbol{b'}^l \right) \quad (2)$$

where $l$ represents the number of hidden layers, $\sigma$ and $\sigma'$ denote the nonlinear activation functions, $\boldsymbol{W}$ and $\boldsymbol{W'}$ are weight matrices, $\boldsymbol{b}$ and $\boldsymbol{b'}$ represent the bias vectors, and $\theta$ denotes the learnable model parameters $\{\boldsymbol{W}, \boldsymbol{b}, \boldsymbol{W'}, \boldsymbol{b'}\}$. $f_\theta^0(.) = \boldsymbol{x}$, and $g_\theta^0(.) = f_\theta^l(.)$.

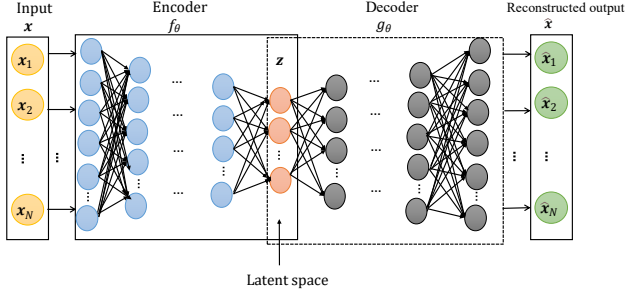

Fig. 1. Structure of a standard autoencoder composed of two nonlinear mappings (fully connected feedforward neural networks) namely the encoder and the decoder.

The training objective of the autoencoder is to minimize the reconstruction error between the output $\hat{x}$ and the input $x$, referred as the loss function $\mathcal{L}(\theta)$, typically the mean square error (MSE), expressed as:

$$\mathcal{L}(\theta) = \sum \|x - \hat{x}\|^2 \quad (3)$$

AE has been widely used for anomaly detection by adopting the reconstruction error as anomaly score. It is trained with only normal data representing the normal behavior. After training, AE will reconstruct the normal instances very well, while it will fail to reproduce the anomalous observations by yielding high reconstruction errors. The process of the classification of an instance as anomalous/normal is shown in Alg. 1.

---

Algorithm 1: Autoencoder based anomaly detection

**Input**: Normal dataset $x$, anomalous dataset $x^{(i)}$ $i = 1, \dots, N$, threshold $\theta$
**Output**: reconstruction error $\|x - \hat{x}\|$
1: train an autoencoder given the normal data $x$
2: **for** $i = 1$ to $N$ **do**
3: $\quad reconstruction\ error\ (i) = \|x^{(i)} - g(f(x^{(i)}))\|$
4: $\quad$ **if** $reconstruction\ error\ (i) > \theta$ **then**
5: $\quad\quad x^{(i)}$ is anomalous
6: $\quad$ **else**
7: $\quad\quad x^{(i)}$ is normal
8: $\quad$ **end if**
9: **end for**

---

### 2.3 Bidirectional Gated Recurrent unit (BiGRU)

The Gated Recurrent Unit (GRU) recently proposed to solve the gradient vanishing problem [14], is an improved version of standard recurrent neural networks (RNNs), used to process sequential data and to capture long-term dependencies. The typical structure of GRU shown in Fig. 2, contains two gates namely reset and update gates, controlling the flow of the information. The update gate regulates the information that flows into the memory, while the reset gate controls the information flowing out of the memory.
The GRU cell is updated at each time step $t$ by applying the following equations:

$$z_t = \sigma(W_z x_t + W_z h_{t-1} + b_z) \quad (4)$$
$$r_t = \sigma(W_r x_t + W_r h_{t-1} + b_r) \quad (5)$$
$$\widetilde{h_t} = \tanh(W_h x_t + W_h(r_t \circ h_{t-1}) + b_h) \quad (6)$$
$$h_t = z_t \circ h_{t-1} + (1 - z_t) \circ \widetilde{h_t} \quad (7)$$

where $z$ denotes the update gate, $r$ represents the reset gate, $x$ is the input vector, $h$ is the output vector, $W$ and $b$ represent the weight matrix and the bias vector respectively. $\sigma$ is the gate activation function and tanh represents the output activation function. The "∘" operator represents the Hadamard product.

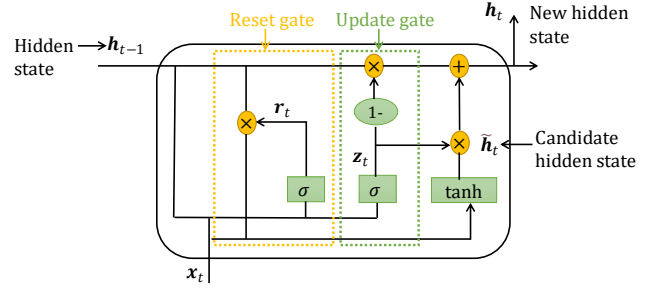

Fig. 2. Structure of the gated recurrent unit (GRU) cell.

BiGRU is an extension of GRU that helps to improve the performance of the model. It consists of two GRUs: one forward GRU model that takes the input in a forward direction, and one backward GRU model that learns the reversed input. The output $y_t$ of the model is generated by combining the forward output $\vec{h}_t$ and backward output $\overleftarrow{h}_t$ as described by the following equations:

$$\vec{h}_t = GRU(x_t, \vec{h}_{t-1}) \quad (8)$$
$$\overleftarrow{h}_t = GRU(x_t, \overleftarrow{h}_{t-1}) \quad (9)$$
$$y_t = \vec{h}_t \oplus \overleftarrow{h}_t, \quad (10)$$

where $\oplus$ denotes an element-wise sum.

### 2.4 Multi-task learning

Multi-task learning is a subfield of ML aiming at improving the overall performance of multiple tasks by learning them jointly while sharing knowledge across them. It has been widely adopted in various fields ranging from natural language processing to computer vision. The multi-task learning approaches can be generally categorized into two architectures: hard- and soft parameter sharing. The hard parameter sharing method is performed by sharing the hidden layers with the different tasks (completely sharing the weights and the parameters between all tasks) while preserving task-specific output layers learnt independently by each task. Whereas for the soft parameter sharing approach, a model with its own parameters is learnt for each task and the distance between the parameters of the model is then regularized to encourage similarities among related parameters [3].

### 3. PROPOSED APPROACH

As shown in Fig. 3, the proposed framework for fiber monitoring can be broken into five main stages: (1) optical fiber network monitoring and data collection, (2) data processing, (3) fiber anomaly detection, (4) fiber fault diagnosis and localization, (5) mitigation and recovery from fiber failures. The optical fibers deployed in the network infrastructure are periodically monitored using OTDR (i.e fiber monitoring unit). The generated OTDR traces (i.e monitoring data) are sent to the software-defined networking (SDN) controller managing the optical infrastructure. Then, the said data is segmented into fixed length sequences and normalized. Afterwards, the processed data is fed to the ML based anomaly detection model for detecting the fiber anomalies or faults. If a fiber anomaly is detected, a ML model for fault diagnosis and localization is adopted to diagnose the fault and localize it. Based on the identified fault, a set of recovery rules is applied to mitigate such fault. The SDN controller notifies the network operation center in case of failure, which informs the customer about the type of detected fault and its location, and notifies the maintenance and repair service in case of fiber cut.

For this work, we consider the fiber anomalies fiber cut and optical eavesdropping attack as examples of harmful or major fiber faults. Given that the patterns of the aforementioned faults and the anomalies bad splice and dirty connector are similar, particularly under very low SNR conditions, we include them during the training phase of the ML model for fault diagnosis to ensure a reliable fault identification and reduce false alarms.

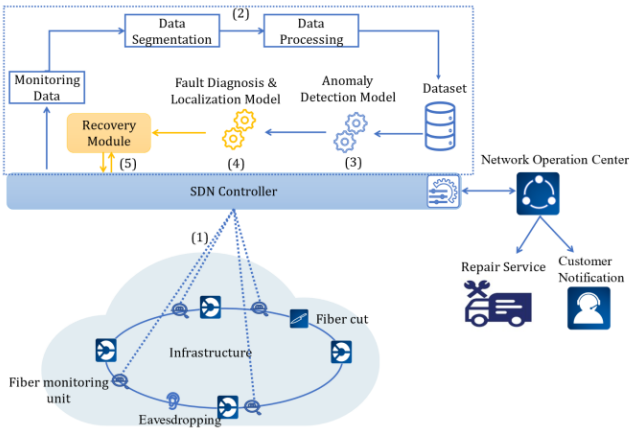

Fig. 3. Overview of the ML-based fiber monitoring process.

### 3.1 Anomaly Detection Model

The autoencoder is trained with normal data only representing the normal behavior in order to learn the distribution characterizing the normal state. After the training of the autoencoder and for the inference phase, the reconstruction error is adopted as an anomaly score to detect any potential fault. A well-trained autoencoder will reconstruct the normal instances very well since they will have the same pattern or distribution as the training data, while it will fail to reproduce the anomalous observations by yielding high reconstruction errors. The process of the classification of an instance as anomalous/normal is illustrated in Algorithm 1. If the computed anomaly score is higher than a set threshold θ, the instance is classified as" anomalous", else it is assigned as" normal". θ is a hyperparameter optimized to ensure high detection accuracy and is adjusted by taking into consideration the degradation and the aging effect of the optical fiber.

The architecture of the proposed autoencoder model for fiber anomaly detection is illustrated in Fig. 4. The model contains an encoder and a decoder sub-model with 4 layers. The encoder takes as an input a 30-length sequence of an OTDR trace $[P_1, P_2, \dots P_{30}]$ representing the attenuation of the fiber along the distance, combined with the sequence's computed SNR ($\gamma$). Including the information about the sequence's SNR during the training phase helps the ML model to learn the behavior of the normal signal pattern for each input SNR level and thereby to boost the performance [3]. The fed input to the encoder is then compressed into low dimensional representation through adopting 2 GRU layers composed of 64 and 32 cells, respectively, which capture the relevant sequential features modelling the normal state under different SNRs. Afterwards, the decoder reconstructs the output, given the compressed representation output of the encoder. The decoder is inversely symmetric to the encoder part. Exponential linear unit (ELU) is selected as an activation function for each hidden layer of the model. The cost function is set to the mean square error (MSE), which is adjusted by using the Adam optimizer.

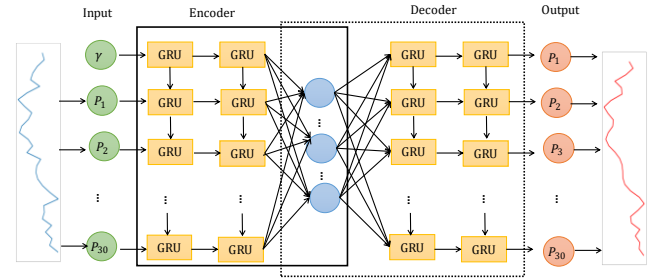

Fig. 4. Structure of the proposed model for fiber anomaly detection.

### 3.2 Fault Diagnosis and Localization Model

As the tasks, namely the fault diagnosis $T_1$, and the fault position estimation $T_2$ can benefit from feature space sharing, the proposed model for fault diagnosis and localization is a multitask learning framework with hard parameter sharing adopted to learn the tasks simultaneously in order to enhance the generalization capability. The architecture of the proposed framework is composed of shared hidden layers distributing the knowledge across the tasks $T_1$ and $T_2$ followed by a specific task layer. The shared hidden layers adopt a combination of bidirectional gated recurrent unit (BiGRU) network and attention mechanisms. The BiGRU is used to capture the sequential features characterizing each fault's pattern, whereas the attention mechanisms help the model to concentrate more on the relevant features in order to improve the fault diagnosis and localization accuracy. As shown in Fig. 5, the input (i.e., the abnormal sample detected by the GRU-based autoencoder) is firstly fed to 2 BiGRU layers composed of 64 and 32 cells, respectively, to learn the relevant sequential features $[h_1, h_2 \dots h_{31}]$. Then, the attention layer assigns to each extracted feature hi a weight (i.e., attention score) αi, which is calculated as follows:

$$e_i = \tanh(W_h\, h_i) \quad (11)$$
$$\alpha_i = \text{softmax}(W^T e_i) \quad (12)$$

where $W_h$, $W$ denote weight matrices. The softmax is used to normalize $\alpha_i$ and to ensure that $\alpha_i \geq 0$, and $\sum_i \alpha_i = 1$.

The different computed weights $\alpha_i$ are aggregated to obtain a weighted feature vector (i.e., attention context vector) $c$, which captures the relevant information to improve the performance of the neural network model. $c$ is computed as follows:

$$c = \sum_i \alpha_i\, h_i \quad (13)$$

Afterwards, $c$ is transferred to two task-specific layers dedicated to solving the tasks of fault diagnosis ($T_1$) and fault localization ($T_2$) respectively, by leveraging the knowledge extracted by the attention-based BiGRU shared layers. The model is trained by minimizing the loss function formulated as:

$$\mathcal{L}_{total} = \lambda_1 l_{T_1} + \lambda_2 l_{T_2}, \quad (14)$$

where $L_{T_1}$ and $L_{T_2}$ denote the loss of $T_1$ and $T_2$, and the first one is the cross-entropy loss whereas the second one is the regression loss (MSE). The loss weights $\lambda_1$ and $\lambda_2$ are hyperparameters to be tuned.

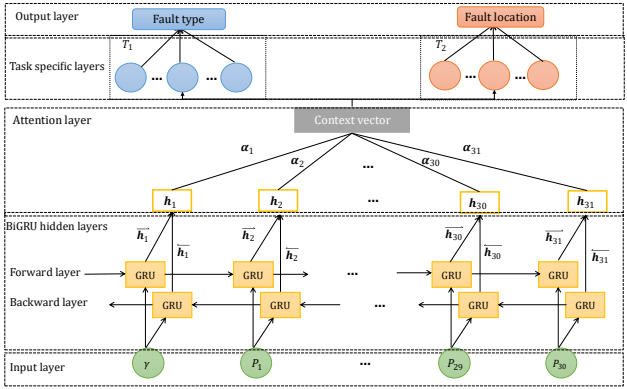

Fig. 5. Structure of the proposed attention-based bidirectional gated recurrent unit model for fiber fault diagnosis and localization.

## 4. VALIDATION OF THE PROPOSED APPROACH

### 4.1 Experimental Data

To validate the proposed approach, the experimental setup shown in Fig. 6 is conducted. The setup is used to record OTDR traces incorporating different types of fiber faults namely fiber cut, fiber eavesdropping (fiber tapping), dirty connector and bad splice. To reproduce a real passive optical network environment, 4 couplers are employed. Optical components like connectors, variable optical attenuator (VOA) and reflector are utilized to model normal events in the fiber optic link. To vary the fiber bending pattern and thereby to enhance the generalizability capability of the ML model, the bend radius of the clip-on coupler is ranged from 2.5 mm to 10 mm. Different bad splices with dissimilar losses are performed to create a varying bad splicing fault pattern. The OTDR configuration parameters namely the pulse width, the wavelength and the sampling time are set to 10 ns, 1650 nm and 1 ns, respectively. From 62 up to 65,000 OTDR records are collected and averaged. Figure 7 shows an example of a recorded OTDR trace incorporating the different faults, whereas Fig. 8 illustrates the patterns of the investigated faults.

### 4.2 Data Preprocessing

The generated OTDR traces are segmented into sequences of length 30 and normalized. For each sequence, γ is computed and assigned. For the training of the GRU-based autoencoder (GRU-AE), only the normal state sequences incorporating either no fault or normal events induced by the optical components, are considered, whereas for testing, both normal samples and faulty sequences incorporating an anomaly, are used. For training GRU-AE, a dataset of 47,904 samples is built and split into a training (70%), and a test dataset (30%).

For training the attention based BiGRU model, we consider only the faulty sequences. For each sequence, the fault type (fiber eavesdropping, bad splice, fiber cut, dirty connector), the fault position defined as the index within the sequence, are assigned. A dataset of 61,849 samples is used for training the fault diagnosis and localization ML model. The said data is divided into a training (60%), a validation (20%) and a test dataset (20%).

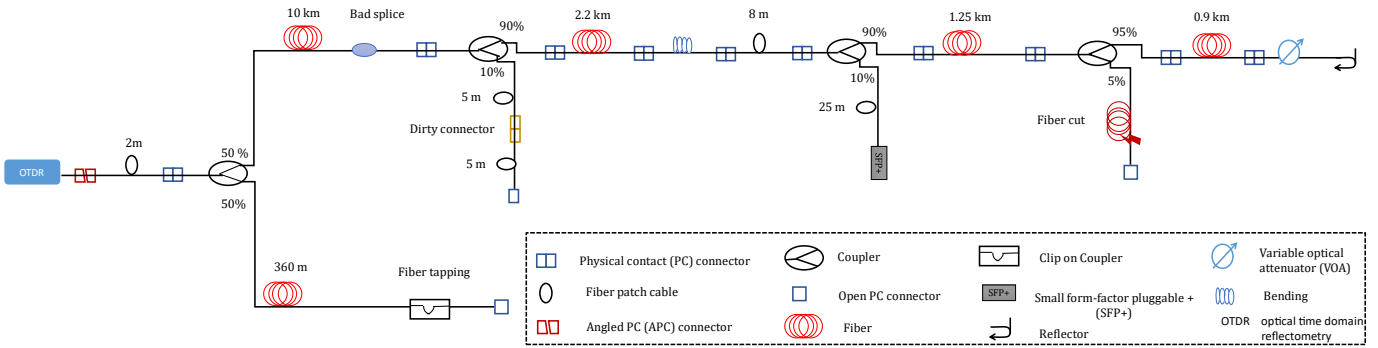

Fig. 6. Experimental setup for generating OTDR data containing different faults induced at different locations in an optical network.

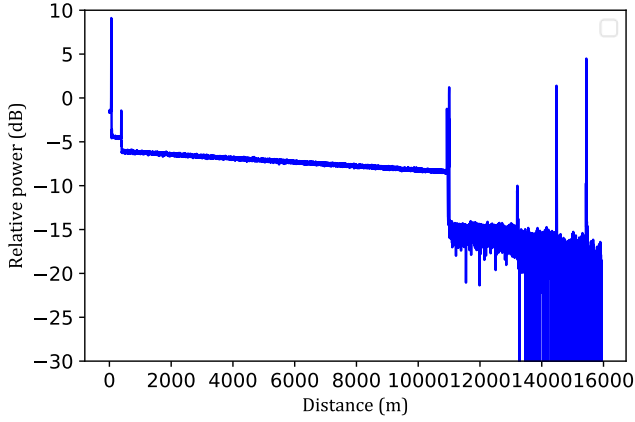

Fig. 7. Example of OTDR trace generated using the experimental setup shown in Fig. 6.

### 4.3 Performance Assessment

#### 4.3.1 Evaluation Metrics

The fault detection is modeled as a binary classification, whereby we distinguish among sequences with label *"1: fault"* (i.e., "positive") or *"0: normal"* (i.e., "negative"). We consider:

- *true positives (TP)*: Number of sequences of type "1" correctly classified with label "1";
- *true negatives (TN)*: Number of sequences of type "0" correctly classified with label "0";
- *false positives (FP)*: Number of sequences of type "0" misclassified with label "1";
- *false negatives (FN)*: Number of sequences of type "1" misclassified with label "0".

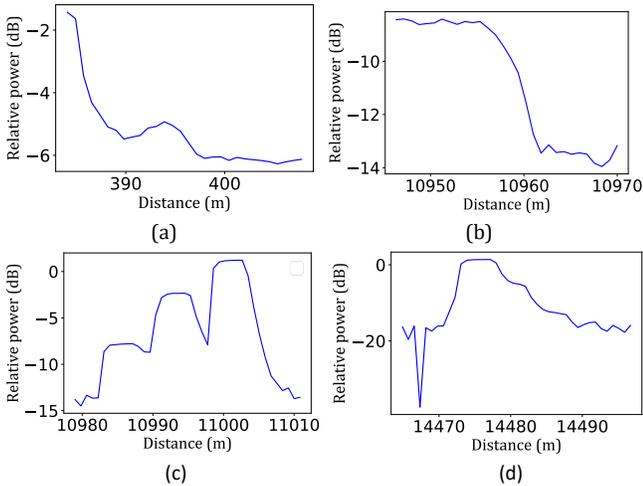

Fig. 8. Patterns of the faults: (a) fiber tapping, (b) bad splice, (c) dirty connector (the second peak, whereas the first and last peaks are induced due to PC connector and open PC connector respectively), (d) fiber cut.

To assess the detection capability, the following metrics are adopted:

- Precision (P) quantifies the relevance of the predictions made by the ML model. It is expressed as:

$$P = \frac{TP}{TP + FP}$$

- Recall (R) provides the total relevant results correctly classified by the ML model. It is formulated as:

$$R = \frac{TP}{TP + FN}$$

- F1 score is the harmonic mean of the precision and recall, calculated as:

$$F1 = 2\frac{P\,R}{P + R}$$

#### 4.3.2 Fault Detection Capability

The anomaly detection capability of GRU-AE is optimized by selecting an optimal threshold θ. Figure 9 shows the precision, the recall, and the F1 score curves as function of θ. If the selected threshold is too low, many faults will be classified as normal, leading to a higher false positive ratio. Whereas if the chosen threshold is too high, many "normal" sequences will be classified as "faulty", resulting in a higher false negative ratio. Therefore, the optimal threshold that ensures the best precision and recall tradeoff (i.e., maximizing the F1 score) is chosen. For the optimally selected threshold of 0.008, the precision, the recall, and the F1 scores are 96.9%, 96.86%, and 96.86%, respectively.

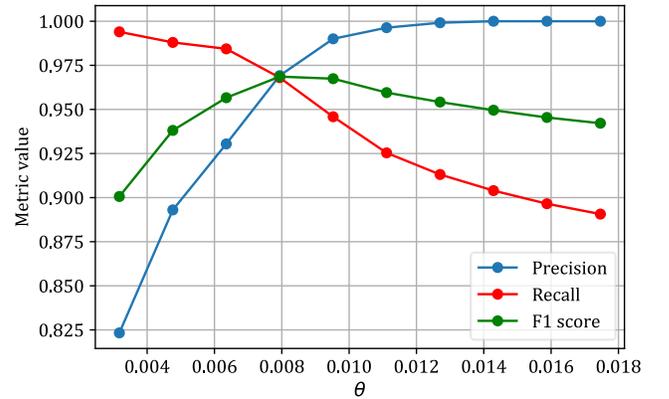

Fig. 9. The optimal threshold selection based on the precision, recall and F1 scores yielded by GRU-AE.

The receiver operating characteristic (ROC) curve, illustrating the performance of the model at different threshold settings, shown in Fig. 10, proves that GRU-AE can distinguish very well between the normal and faulty classes by achieving a high area under the curve (AUC) (i.e., measuring the degree of separability between the classes) of 0.98.

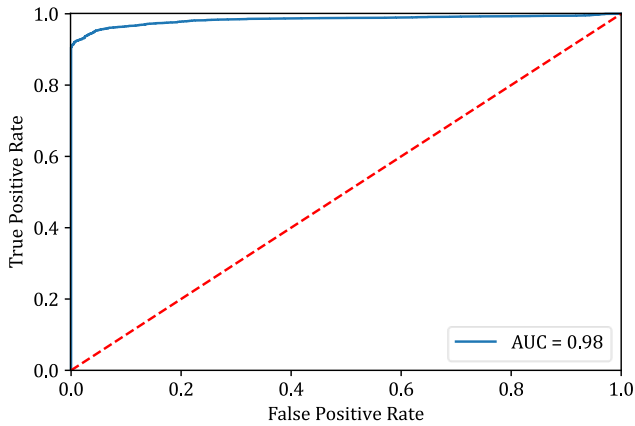

Fig. 10. The receiver operating characteristic curve of GRU-AE.

### 4.3.3 Optimization of GRU-AE

Network architectures with various sizes are evaluated to select the optimum one, achieving the best performance while ensuring a moderate complexity. The impact of the parameters, namely "depth" (i.e., the number of hidden layers), "width" (i.e., number of cells for each hidden layer), and the activation function, on the performance of GRU-AE in terms of reconstruction error is investigated. The number of hidden layers with either 32 or 64 cells is firstly varied from 2 to 8. Different activation functions for the hidden layers, namely rectified Linear Unit (ReLU), leaky ReLU, scaled exponential linear unit (SELU), and ELU, are analyzed. Figure 11 shows the output of each activation function for a given input. Several combinations of number of cells for each hidden layer of the encoder network are tested.

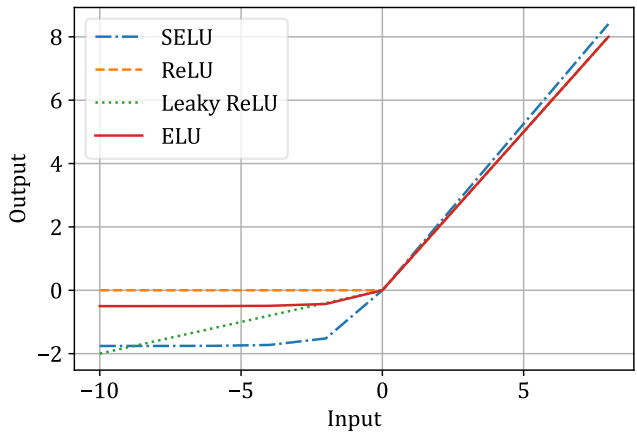

Fig. 11. The output of different activation functions.

As shown in Fig. 12 (a), the reconstruction error of GRU-AE shows a decreasing trend with the increase of the depth for a different number of cells per hidden layer, before reaching the optimum depth of 4. Increasing the depth helps the GRU-AE model to capture more features modelling the normal behavior. However, widening the layers higher than 6 can lead to overfitting and thus reduces the performance of the GRU-AE. As illustrated in Fig. 12 (b), adopting ELU as activation function in the hidden layers achieves the smallest reconstruction error compared to the other functions. Figure 12 (c) confirms that setting the number of cells for each layer of the encoder to 64 yields the best performance (lowest reconstruction error).

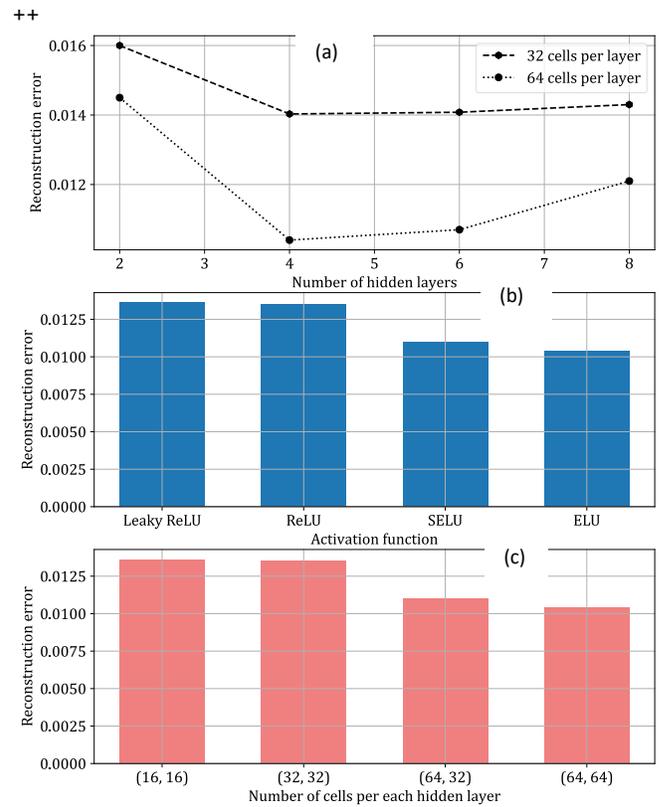

Fig. 12. Optimization of the GRU-AE model: (a) reconstruction errors with different depths for different cells per layer, (b) for different activation functions in the hidden layers, (c) for several number of cells per hidden layer of the encoder.

### 4.3.4 Comparison of GRU-AE with other ML models

The GRU-AE model is compared to other unsupervised anomaly detection methods namely isolation forest (IF), local outlier factor (LOF), and one-class support vector machine (OCSVM), in terms of F1 score and the area under the precision recall curve (AUPRC) metrics. IF isolates the outliers in the data by performing a random partition on the data observations and then computing the split value between the maximum and minimum of the chosen instance. The path length defined as the number of splits required to isolate each observation is adopted as anomaly score. LOF computes the anomaly score by measuring the local deviation of the density of a given instance with respect to its neighbors. OCSVM learns the boundary decision encompassing the normal data in the feature space, and during the inference stage, any sample lying outside that boundary is considered an anomaly. IF, LOF, and OCSVM are trained with normal data, like the GRU-AE model, and tested with unseen data including both normal and abnormal data. The results shown in Table 1 demonstrate that the GRU-AE model outperforms the other ML methods by yielding the highest values of F1 score and AUPRC scores. Compared to the tested ML algorithms GRU-AE provides significant improvements of more than 5.5% and 6.36% in AUPRC and F1 scores, respectively.

TABLE I
COMPARISON OF DIFFERENT ANOMALY DETECTION ML METHODS IN TERMS
OF AREA UNDER THE PRECISION RECALL CURVE (AUPRC) AND F1 SCORE.
THE BEST RESULT IS SHOWN IN BOLD.

| Method | F1 score (%) | AUPRC (%) |
|---|---|---|
| OCSVM | 49.8 | 94.1 |
| IF | 86.9 | 71.5 |
| LOF | 90.5 | 61.7 |
| GRU-AE | **96.86** | **99.6** |

**4.3.5 Fault Diagnosis Capability**

The confusion matrix shown in Fig. 13, proves that the attention based BiGRU model (A-BiGRU) diagnoses the different faults with an accuracy higher than 97%, and accurately distinguishes the physical fiber attack by achieving an accuracy of 98%. As for low SNR sequences, the patterns of eavesdropping and bad splice faults look similar, the ML model mis-classified a little these classes. The same applies for dirty connector and fiber cut patterns under low SNR conditions leading to low misclassification rates.

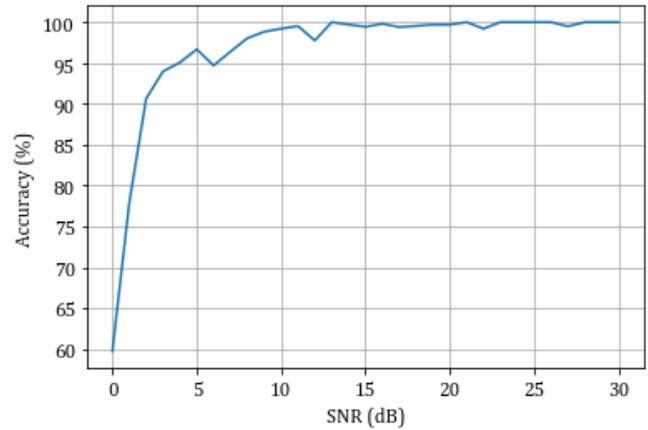

Fig. 14. The diagnosis accuracy of the A-BiGRU model.

Fig. 13. The confusion matrix of A-BiGRU model.

Figure 14 shows the effects of SNR on the diagnosis accuracy of A-BiGRU. The accuracy increases with SNR. For SNR values higher than 10 dB, the accuracy is approaching 100%. For an SNR lower than 2 dB, the accuracy is worse as it is very difficult to differentiate the different faults mainly due to the noise that adversely impacts the patterns of the faults which might look similar under very low SNR levels. For input sequences of SNR higher than 2dB, A-BiGRU could discriminate the different types of faults with a good accuracy higher than 90.5%.

The feature learning ability of A-BiGRU under very low SNR conditions (SNR ≤ 5 dB) for solving the task $T_1$ is visually investigated using the t-distributed stochastic neighbor embedding (t-SNE) technique [15]. Figure 15 shows that first the learned features under SNR levels lower than 1 dB are of very poor separability and A-BiGRU misclassifies most of the faults as fiber cut mainly due to the similarity of the different fault patterns under those SNR conditions because of the high noise overwhelming the patterns, second the extracted features become more and more discriminative with the increase of the SNR, and third A-BiGRU can learn effective features for accurate fault diagnosis even for an SNR condition higher than 2 dB.

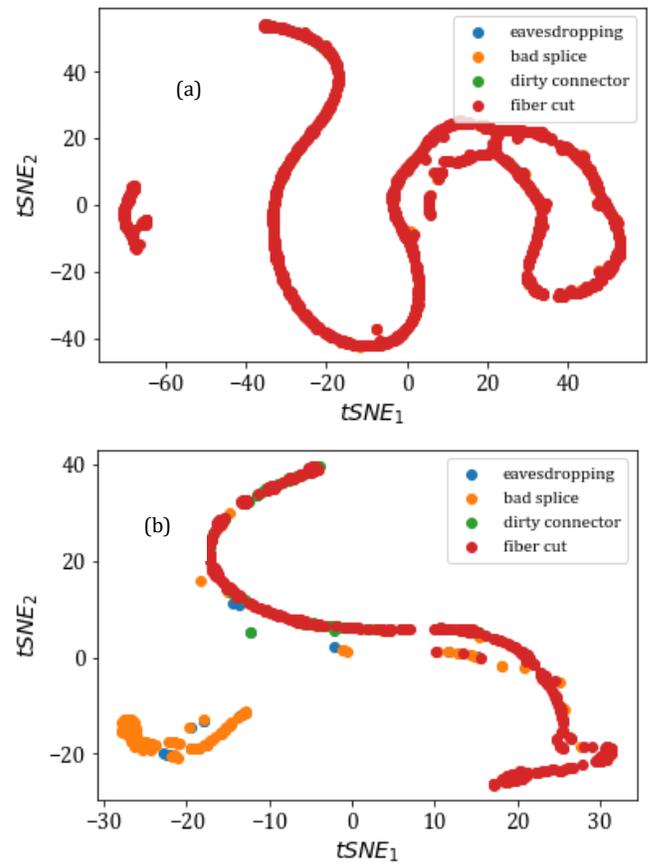

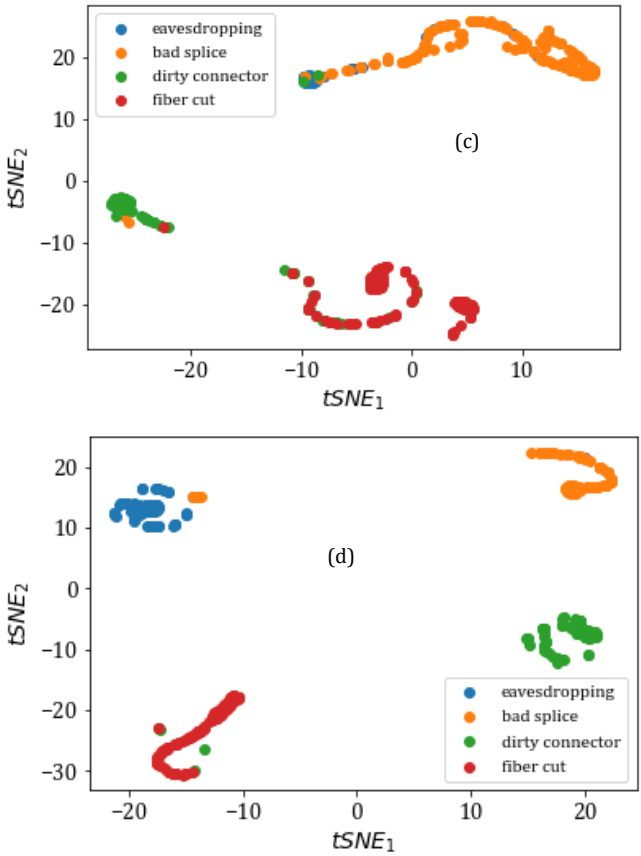

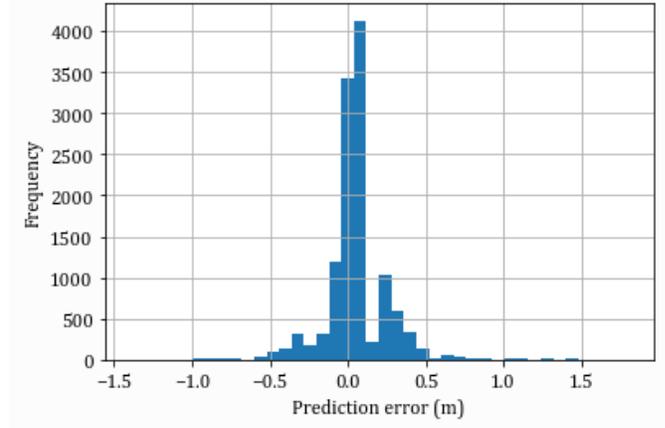

Fig. 16. Histogram of position prediction errors yielded by A-BiGRU model.

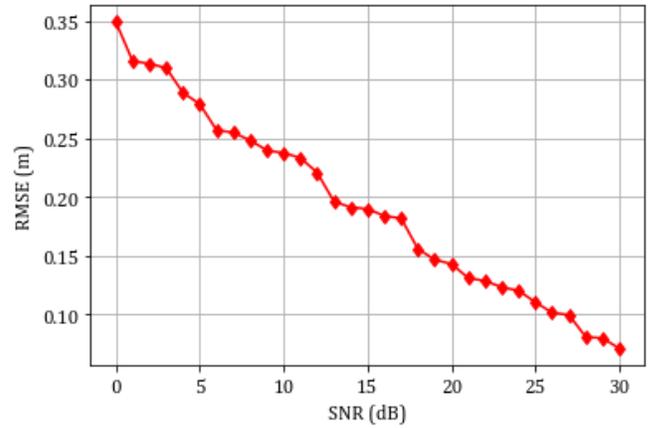

Fig. 15. Visualization of the feature learning under low SNR conditions: (a) 0 dB SNR, (b) 1 dB SNR, (c) 2 dB SNR, (d) 5 dB SNR.

**4.3.6 Fault Localization Capability**

To assess the fault localization accuracy of the A-BiGRU model, two evaluations metrics namely the prediction error representing the difference between the predicted and actual position of the fault, and the root mean square error (RMSE) are adopted. Figure 16 shows that A-BiGRU achieves very small prediction errors with a mean of 0.05 m and a standard deviation of 0.2, which proves that the ML model accurately localizes the faults. We analyzed the fault localization estimation performance of A-BiGRU as function of SNR. As depicted in Fig. 17, A-BiGRU accurately localizes the faults by achieving an average root mean square error (RMSE) of 0.19 m, and that the RMSE decreases with SNR. For lower SNR values (SNR ≤ 10 dB), the RMSE can be higher than 0.35 m, whereas for SNR values higher than 13 dB, it is less than 0.2 m and it could be further reduced up to less than 0.1 m for SNR values higher than 27 dB.

Fig. 17: Fault position estimation error (RMSE) for the ML model.

**4.3.7 Comparison of BiGRU with other existing ML approaches**

The BiGRU is compared to two baseline ML models recently proposed for fiber fault diagnosis and localization, namely BiLSTM-CNN [6] and BiLSTM [7]. For the sake of a fair comparison, BiLSTM-CNN and BiLSTM are retrained with the same training data as BiGRU, and we adjusted their structures for solving only the two learning tasks $T_1$ and $T_2$. For the architectures of the models, BiLSTM is composed of one BiLSTM layer with 32 cells followed by task-specific layers composed of 16 and 20 neurons, respectively. In contrast the BiLSTM-CNN model consists of one BiLSTM layer with 32 cells followed by CNN layers containing mainly one convolutional layer having 32 filters with the max pooling layer succeeded by a dropout layer, and two task-specific layers composed of 16 and 20 neurons, respectively. The length of the input sequence of both models is set to 30. We compare the different models by adopting as evaluation metrics the average diagnostic accuracy to assess the fault diagnosis capability, and the average RMSE to evaluate the fault localization performance. Figure 18 (a) proves that the proposed model outperforms the existing methods by achieving an improvement of 8.8% in accuracy due to the inclusion of the SNR during the training phase and the adoption of the attention mechanisms to capture the relevant features and thereby boost the performance. Figure 18 (b) shows that the proposed method achieves the lowest RMSE value compared to the other models.

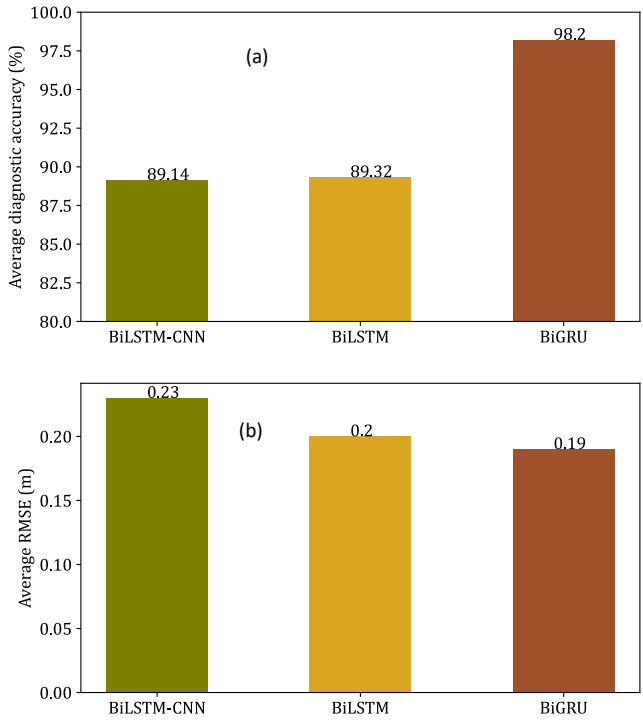

Fig. 19: Comparison between the proposed model and existing ML methods in terms of: (a) average diagnostic accuracy and (b) average RMSE.

The comparison of the results of computational inference time between the proposed model and the existing methods are shown in Table 2. As it can be seen, the proposed model consumes slightly more time than the existing methods due to its deeper architecture.

TABLE II
COMPUTATIONAL TIME OF POPOSED MODEL AND EXISTING METHODS.
THE BEST RESULT IS SHOWN IN BOLD.

| Method | Inference time (12,370 samples) |
| --- | --- |
| BiLSTM | $1.06 \pm 0.03\ s$ |
| BiLSTM-CNN | $1.18 \pm 0.14\ s$ |
| BiGRU | $2.1 \pm 0.13\ s$ |

#### 4.3.8 Integrated learning approach GRU-AE- BiGRU

The performance of the integrated approach combining GRU-AE and BiGRU models, called model A, is compared to a BiGRU model trained to discriminate between the normal state and the different types of faults (omitting the use of the autoencoder), denoted by model B, in terms of average accuracy metric. The model B is trained with data including both normal and faulty data modelling the different fault classes. Figure 19 shows the confusion matrix obtained by model B. As it can be noticed, it is hard for the model to distinguish between the normal class and the other faults namely bad splices and eavesdropping events due to the similarity of their patterns under low SNR conditions.

Fig. 19. The confusion matrix of model B.

Table 3 proves that model A outperforms model B in terms of accuracy with an improvement of 5.1%, which demonstrates the importance of adopting the autoencoder to discriminate the normal class from the faulty classes and thereby enhances the fault diagnosis capability of BiGRU and reduces the false alarm rate.

TABLE III
COMPARISON OF ML MODELS IN TERMS OF AVERAGE ACCURACY. THE BEST RESULT IS SHOWN IN BOLD.

| Method | Average accuracy (%) |
| --- | --- |
| Model A (GRU-AE + BiGRU) | **96.9** |
| Model B (without GRU-AE) | 91.8 |

#### 4.3.9 Investigation of the Robustness of BiGRU

Given that the BiGRU model is trained with data incorporating the faults induced at fixed locations of the network to assess the robustness capability of the proposed method, we modify the locations of the different faults as shown in Fig. 20, and test the performance of BiGRU given the new data generated using the new experimental setup.

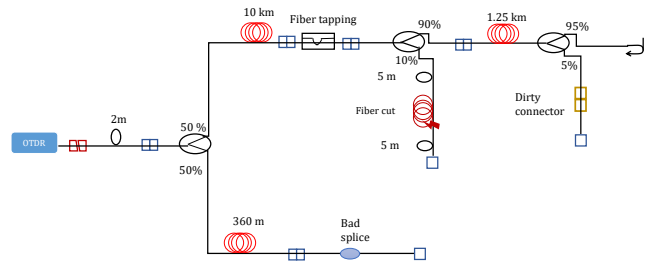

Fig. 20: Modified experimental setup for testing the robustness of the BiGRU model. For a legend of symbols, please refer to Fig. 6.

Tested with that data, the BiGRU achieves a good fault diagnosis capability by yielding an accuracy of 96.8%, proving that the ML model effectively learns the different types of faults, and thus is capable of diagnosing them at different locations of the network.

## 5. CONCLUSION

A ML-based approach for fiber fault detection, identification and localization is proposed. The presented framework includes an autoencoder model to detect the fiber anomalies and an attention based bidirectional gated recurrent unit method to recognize the detected fiber faults and localize them. The effectiveness of the proposed approach is validated using OTDR data incorporating various faults including fiber cuts and optical eavesdropping attacks. The experimental results proved that the presented framework achieves a good fault detection and diagnosis capability and a high localization accuracy. Our experiments show that ML techniques can enhance the performance of the anomaly detection and fault diagnosis and localization in fiber monitoring, and thereby minimize the false positive alarms, saving time and maintenance costs. In our future work, we plan to  which is usually operated in a very complex, sophisticated, intelligent, and autonomous environment.

**Acknowledgements.** This work has been performed in the framework of the CELTIC-NEXT project AI-NET-PROTECT (Project ID C2019/3-4), and it is partly funded by the German Federal Ministry of Education and Research (FKZ16KIS1279K).